\documentclass[prl,reprint,twocolumn,amsmath,amssymb,
 aps]{revtex4-2}

\usepackage{natbib}
\usepackage{xspace}
\usepackage{verbatim}
\usepackage{graphicx}
\usepackage{array}
\usepackage[space]{grffile}
\usepackage{latexsym}
\usepackage{amsfonts,amsmath,amssymb}
\usepackage{url}
\usepackage[utf8]{inputenc}
\usepackage{fancyref}
\usepackage {cases}
\usepackage{float} 
\usepackage{savesym}
\savesymbol{tablenum}
\usepackage{siunitx}

\usepackage{appendix}
\usepackage{xcolor}
\newcommand\apjl{{\@eapj@ApJLetters}}

\newcommand{\orcid}[1]{\href{https://orcid.org/#1}{\includegraphics[height=\fontcharht\font`\B]{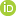}}}

\newcommand{\JHU}{Department of Physics and Astronomy, Johns Hopkins University, Baltimore, MD 21218, USA}
\newcommand{\Salamanca}{Department of Fundamental Physics and IUFFyM, University of Salamanca,\\ Plaza de la Merced S/N E-37008, Salamanca, Spain}
\newcommand{\Oxford}{Department of  Physics, University of Oxford, Keble Road OX1 3RH, Oxford, UK}
\newcommand{\CNRS}{Institut d’Astrophysique, UMR 7095 CNRS, Sorbonne Universit\'e, 98bis Blvd Arago, 75014 Paris, France}

\newcommand{\ISEF}{ISEF Fellowship}

\begin{document}

\title{Neutrino signals from Neutron Star implosions to Black Holes}


\author{Yossef Zenati~\orcid{0000-0002-0632-8897}}
 \email{yzenati1@jhu.edu}
\altaffiliation{\ISEF}
\affiliation{\JHU}

\author{Conrado Albertus, M. \'Angeles P\'erez-Garc\'ia}
 \affiliation{\Salamanca}

\author{Joseph Silk}
\affiliation{\Oxford}
\affiliation{\CNRS}
\affiliation{\JHU}

\date{\today}

\begin{abstract}
We calculate the neutrino luminosity 
in an astrophysical scenario where dark matter 
is captured by a neutron star which eventually implodes to form a low mass black hole. The Trojan horse scenario involves the collapse of a neutron star (NS) due to the accumulation of a critical amount of dark matter (DM) during its lifetime. As a result, a central disk forms out of the ejected material with a finite radial extension, density, temperature, and lepton fraction,
producing fainter neutrino luminosities
and colder associated spectra than found in a regular core-collapse supernova. The emitted gravitational wave (GW) signal from the imploding NS should be detectable at ultra-high $\gtrsim 0.1$ GHz frequencies.
\end{abstract}

\keywords{dark matter --- black hole --- neutron star --- implosion --- neutrinos} 

\maketitle

\textit{Introduction.—} Massive stars, larger than about $\sim 8 M_\odot$ have  short life-times and end in core-collapse supernovae (CCSN), leaving behind a high-mass neutron star (HNS) or stellar black hole (BH). One important quantity in this CCSN event is the rate of neutrinos traveling from the inner regions of the core of the massive star and reaching to the stellar surface. Neutrino physics is still under investigation in all areas of astrophysics and fundamental physics itself, including neutrino-matter interactions, energy/momentum/lepton number transport, and flavor conversion \cite{Athar_2022}. Neutrinos can change their flavors during propagation, affecting the final yield and elements formed in the ejecta or in the outflow both in CCSN events and in binary neutron star (BNS) mergers, as known from the recent kilonova (KN) event  AT~2017gfo, compatible with a BNS merger event at 40 Mpc \cite{distance}, see \cite{Metzger19, Nakar20, Radice+20} arising $\sim$~0.5 days after the gravitational wave (GW) emission GW170817. The latter was detected by the LIGO/Virgo collaboration \citep{abbott2017a}, with a simultaneous short gamma-ray burst (GRB) observed \cite{goldstein2017,blackburn} by Fermi and INTEGRAL (GRB~170817A).

More than five decades after the detection of the first pulsar \cite{1968Natu}, basic properties of neutron stars (NSs) such as their masses and radii are still quite uncertain \citep{OzelFPauloF16}. Understanding the features of the M-R curve for NS is key to setting constraints on the high-density part of the matter equation of state (EoS) and the resulting massive objects left in mergers of NS, see a novel study of NS mass distributions based on GW analysis in \cite{ChatziioannouFarrPRD20}. Regarding composition, the interiors of these objects remain poorly known. Besides the hadronic content that should populate the ultradense NS core, the actual degrees of freedom are unknown. Whether it is a pure nucleonic or a hybrid quark object is undetermined, while consequences and signatures of the possible phase transition have now been thoroughly studied in the literature, see \cite{Baym_2018} for a review. 
Analogously, from seminal papers \cite{1985ApJ...296..679P, Gould:1987ir} there is now a vast literature about the possibility that matter of a different nature, yet unknown, but with feeble interactions with Standard Model (SM) particles, can clump in the Universe and populate astrophysical bodies.

The evidence for dark matter (DM) is currently overwhelming, ranging from galaxy rotation curves, gravitational lensing, Cosmic Microwave Background (CMB) and more \cite{Bertone_2018}. Although modified theories of gravity attempt to explain most observations, to date a more attractive
and accepted explanation is that DM constitutes most of the matter content in our Universe under the form of cold, essentially collisionless particles. Concerning the zoo of current DM candidates in the dark sector, there are some candidates which are especially popular, such as WIMPs \cite{bertone2005}, axions (or axion-like particles, i.e. ALPs) \cite{Choi_2021} and primordial black holes (PBH), see \cite{Carr_2022} for a review.

PBHs are interesting DM candidates generated by inflation and due to tailored primordial fluctuations, and they can explain all of the DM in some mass windows \citep{carr2021}.
Note that from current constraints, there is a possibility that a fraction of DM is in the form of PBH coexisting with additional particle candidates. Determining the optimal PBH mass window remains problematic and is highly constrained from recent GW detections associated with binary BHs in the mass range 10-50 $M_\odot$ and by gravitational microlensing experiments over the mass range $10^{-9}-10^2 M_\odot$.

%

Here we focus on the interface of dense clouds of DM, NS and BH. In dense clouds of DM, NS can accrete a significant amount so that collapse is triggered. The effect is especially strong if the dark matter is asymmetric a.k.a. non self-annihilating \citep{PhysRevLett.113.191301,dasgupta2021, bhattacharya2023}.
An alternative pathway to NS collapse is via capture of a PBH \citep{richards2021} as we detail below. {The question we address here is how can we test the remarkable scenario of induced NS collapse.}

There is wide agreement about how stellar BH are formed via CCSN as the core engine runs out of fuel, no longer being able to stop the gravitational contraction. This fate arises above the NS mass upper limit $\sim 2.1 M_{\odot}$. 

In 2019, it was reported \cite{Abbott_2020} that the GW190814 event was consistent with a binary BH merger where the lightest object $\sim2.6 \rm M_\odot$ was most likely in a mass gap between roughly $2-5 \rm M_\odot$ where boundaries comprise the heaviest NS and the lightest BH that theoretical approaches predict. In addition, a recent analysis using Hubble Space Telescope archival data and densely sampled light curves from ground-based microlensing surveys has spotted one compact object, \texttt{OB110462}, that is  an isolated BH with an inferred lens mass of about $\sim 1.6-4.4 \rm M_\odot$ \cite{Lam_2022}. 

Yet as an alternative scenario, it has been pointed out that low mass BH may be forming from the induced collapse of a NS due to  the accretion of DM to form a fermionic or bosonic self gravitating BH \citep{PhysRevD.40.3221, PhysRevD.85.023519} or from self-annihilating DM by a Trojan Horse mechanism nucleating quark bubble instabilities \citep{Perez-Garcia+10, Herrero+19}. While the former scenario involves critical numbers (mass) of DM particles being accreted over the stellar lifetime, there is a possibility that ordinary NS with masses below the maximum mass value currently at $\sim 2.1 \rm M_\odot$ plausibly follow this fate.

In such a case, the induced implosion of the NS has been shown to produce multi-messenger emission, mostly electromagnetic transients \cite{P_rez_Garc_a_2013} or cosmic rays \cite{Kotera_2013,_ngeles_P_rez_Garc_a_2014}. These transients are not accompanied by significant gravitational radiation or neutrinos, allowing such events to be differentiated from compact object mergers occurring within the distance sensitivity limits of GW observatories, see \cite{FullerPhysRevLett.119.061101, Bramante_2016}. 


\begin{figure}[t!]
    \begin{center}
        \includegraphics[width=0.99\linewidth]{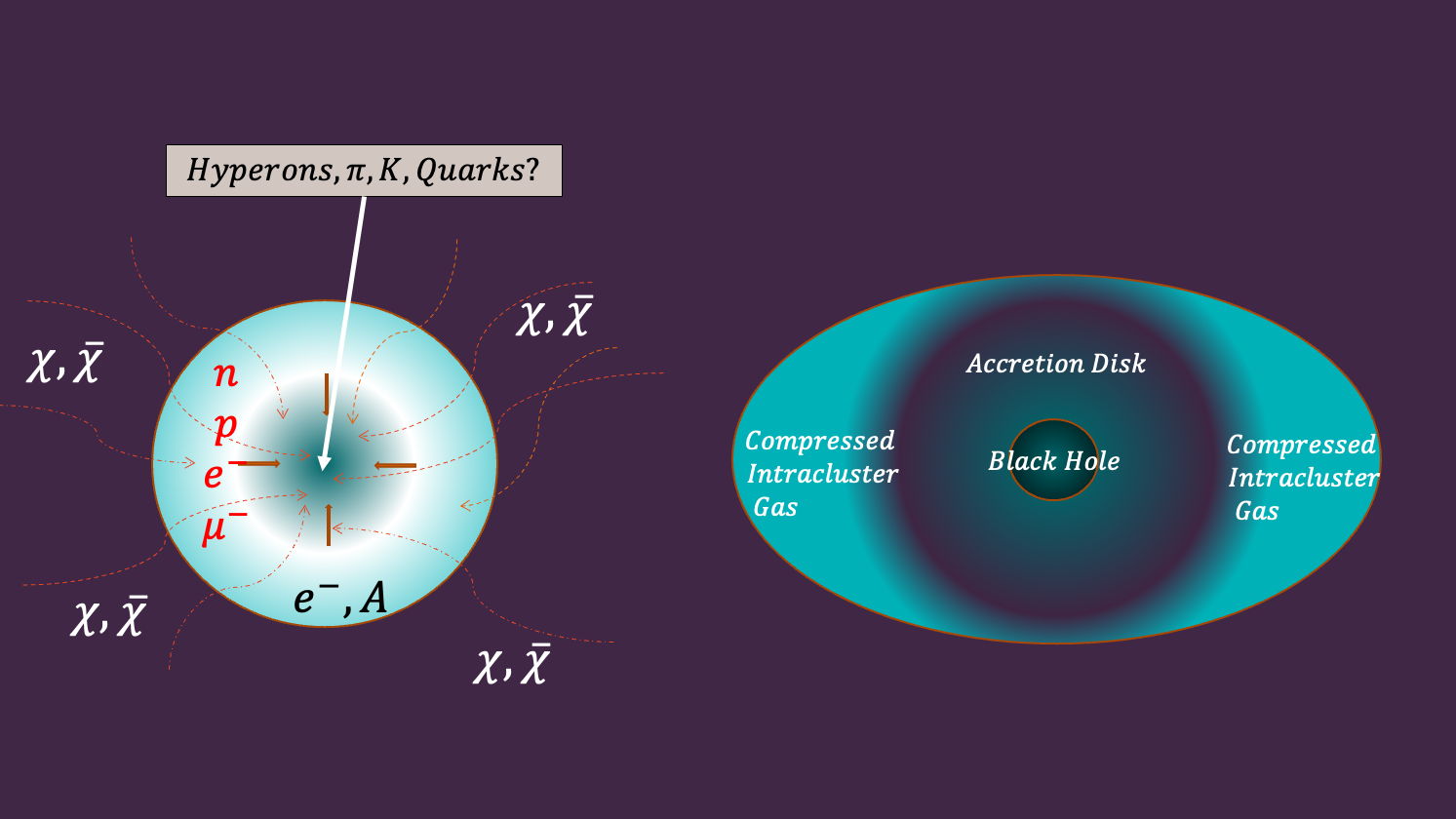}
            \caption{Cartoon showing (left) the NS implosion due to a critical accumulation of an inner core of DM, from a generic massive candidate $\chi$. Following this event, a BH with an accretion disk (right) of mostly ordinary matter and radial extension $R_{disk}$ may be formed. A BH face-on view is shown.}
        \label{fig:spinning}
    \end{center}
\end{figure}



\textit {Dark Matter inside an isolated NS.—} In a regular NS with ordinary matter EoS, the static or rotating solution of the stellar structure equations in the general relativistic framework predicts the mass-radius (M-R) curve where compact stars can exist in stable configurations \citep{Glendenning:1997wn} up to a maximum mass. The actual value of the maximum NS mass is an important observable, as it constrains the poorly known matter EoS at high densities. For isolated NS, it has been found that for $M < 2.2 M_\odot$, it is not likely that a BH is formed, according to the usual stellar gravitational collapse scenario. However, as cited above, several novel mechanisms involving DM have been studied that could distort, in principle, stellar stability. The perturbation mechanism will depend on the actual DM particle candidate and its interaction with Standard Model particles or fields in the hybrid object. As an example, let us consider the presence of asymmetric dark matter (ADM) inside the NS. As studied by several authors, DM in the form of massive particle candidates could be gravitationally captured and accumulated in the NS core \citep{bell2020} in sufficient numbers to trigger the gravitational collapse once a dark critical mass (or particle number $N_{\chi,crit}$) is reached \cite{Kouvaris08PRD, Angeles+22PRD}.

The bosonic or fermionic nature assumed for the dark sector would not change the basic picture as presented here. 
In the case of self-annihilating DM, the {\it Trojan horse} mechanism \cite{Perez-Garcia+10, Herrero_2019} of efficient spark injection producing quark nucleation in the core of the NS may lead to long-hypothesized quark star formation and subsequently a BH. This transition to deconfine the quark content is unlikely in the standard scenario since it requires that the star slows down its rotation so that its central density rises and crosses the critical density of the phase transition. Thus, intermediate mass stars are more likely to have quark seeding in their life-times at birth while those of lower mass remain regular NS throughout their life-times. We start from a treatment where interactions are essentially gravitational with a simplified two-fluid formalism. An admixed star with ordinary matter (fluid 1) and generic DM candidates $\chi$ (fluid 2) display a M-R curve that shows different maximum masses allowed for baryonic and dark overlapping distributions as obtained from solving the structure equations. In the static approximation, the spherically symmetric equations for both fluids read
\begin{equation}
\frac{d p_1}{d r}=-\frac{G M \rho_1}{r^2} \frac{\left(1+\frac{p_1}{\rho_1 c^2}\right)\left(1+\frac{4 \pi}{M c^2} r^3\left(p_1+p_2\right)\right)}{\left(1-\frac{2 M G}{c^2 r}\right)}
, \label{TOV}
\end{equation}
\begin{equation}
\frac{d p_2}{d r}=-\frac{G M \rho_1}{r^2} \frac{\left(1+\frac{p_2}{\rho_2 c^2}\right)\left(1+\frac{4 \pi}{M c^2} r^3\left(p_1+p_2\right)\right)}{\left(1-\frac{2 M G}{c^2 r}\right)} 
\end{equation}
and $\frac{d M_1}{d r}=4 \pi r^2 \rho_1(r) \quad
\frac{d M_2}{d r}=4 \pi r^2 \rho_2(r)$ with total mass $M(r)=M_1(r)+M_2(r)$. $p_1(\rho_1)$ and $p_2(\rho_2)$ are the EoS for both fluids. 

It has been found \cite{dasgupta2021} that for particle ADM candidates with mass above a value $m_\chi=10^8$ GeV, 
there are admixed NS solutions that yield the maximum mass  around the threshold $\sim 1 M_\odot$ so that beyond this value, collapse to a BH would follow \citep{bhattacharya2023}. 

Typically, a massive NS with mass $M$ and radius $R$ is born with a very high rotation frequency, $\omega$, near the Keplerian frequency value, $\Omega_{K}$ \citep{1989Natur.340..451S}. Studying the effect of $\Omega$ on a fluid element with angular velocity relative to that of the local inertial frames $\omega(r)$ in a rotating NS shows $\Omega_K \simeq \left[1+\frac{\omega(R)}{\Omega_K} - 2\left(\frac{\omega(R)}{\Omega_K}\right)^2\right]^{-0.5} \left(\frac{G M}{R^3}\right)^{0.5}$, so that using the relevant solution of the Einstein equations, one can approximate $\Omega_{K}\sim 0.65\left(G M/R^3\right)^{0.5}$ \cite{Glendenning2000}. 
In addition, the relation $\omega(R) / \Omega_K=2 I / R^3$ is fulfilled, being $I$ the stellar moment of inertia . We note at this point that although correlation effects in rotating NS regarding different EoS and its effects on the stellar structure are indeed possible, we will not take them into consideration as it is out of the scope of this work. 


\textit{Neutrino emission.—} To model the disk surrounding the BH, we consider electrically neutral matter with electrons ($e^-$), muons ($\mu^-$), protons (p), and neutrons (n) which are to good approximation consistent with the thermodynamical conditions $\rho, T$. We adopt the polytropic EoS of dense matter \texttt{PAL}, \texttt{APR} and \texttt{PPEOS} including the cold pressure and strongly interacting components $(n,p)$ \citep{Prakash+88PRL, Lattimer&Prakash00, Ngo&Shinmura17, Schneider+19}. In this work, we initialize the electron fraction $\rm Y_e$ by the initial condition of neutrino-free beta-equilibrium \mbox{$\mu_{\nu}(\rho_b,Y_e,T)=0$}, where $\mu_{\nu}$ is the neutrino chemical potential. Generally, the GW signal and the energy emitted throughout is highly dependent on the NS EoS. 

The disk surrounding the BH formed from the debris of the NS implosion can be described in terms of an electron fraction or ratio of electron to baryon number densities in the form $Y_e = n_e/n_b$. We will assume baryon number conservation so that proton and neutron number densities fulfill $n_b = n_p + n_n$, and electrical charge neutrality involving $n_e+n_\mu=n_p$ being  $n_\mu$ the muon number density. The specific form for the previous expressions is determined by the degree of lepton and nucleon degeneracy $\eta_{i}=\mu_{i} / T$, i.e. chemical potential (density) and temperature in the system. 

\begin{figure}[h]
    \begin{center}
        \includegraphics[width=1.0\linewidth]{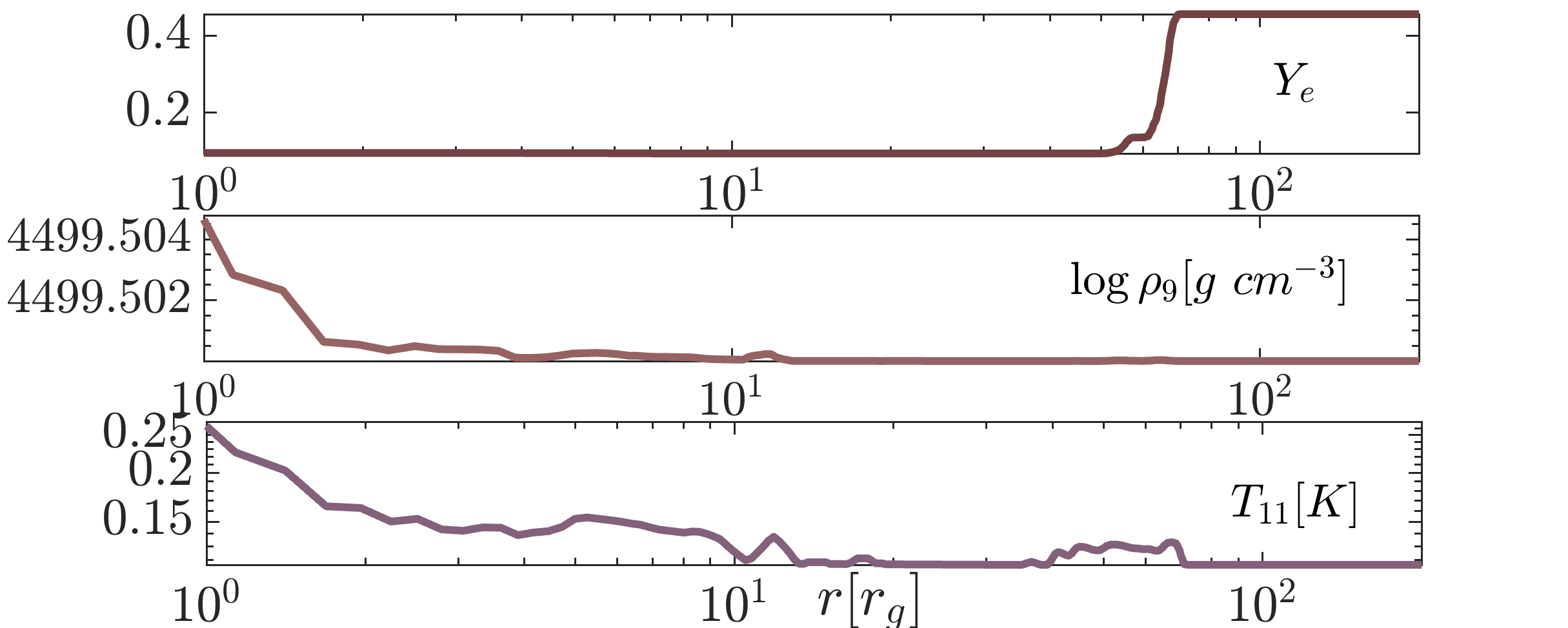}
        \caption{(Top to bottom) Electron fraction, mass density normalized to $10^9$ $\rm g/cm^3$, temperature normalized to $10^{11}$ K as functions of radial coordinate in $r_{g}$ units  at $t = 1\ ms$ after the collapse event.}
        \label{fig:Ye_rho_temp}
    \end{center}
\end{figure}

In Fig. \ref{fig:Ye_rho_temp}, we show some quantities of interest as functions of the radial coordinate for (ordinary) matter beyond the horizon as given from the Schwarzschild BH radius $r_{g}=2GM/c^2$. We show (from top to bottom) electron fraction, $Y_e$, mass density $\rho$ normalized to $10^9$ $\rm g/cm^3$, temperature $T$ normalized to $10^{11}$ K at $t = 1\ ms$ after the collapse event \citep[e.g,][]{Prakash+88PRL,Lattimer&Prakash00} We can see that neutron-rich matter at densities 
 $10^{11-12}$ $\rm g/cm^3$ is present for distances $[1,60]r_g$ to the BH while it is isospin-symmetric beyond distance $ct$. Similarly, T decreases slowly with larger $r$ with an average value around $\sim 1$ MeV.

All the subscripts standing for neutron and electron can be written in terms of two weak  equilibrium reactions $n \leftrightarrow p + e^- + \bar \nu_e$ and $\mu^- \leftrightarrow e^- + \bar \nu_e +  \nu_{\mu}$. Particle populations can be expressed by chemical potential relations $\mu_p = \mu_n - \mu_e$, $\mu_\mu = \mu_e$ for neutrinos escaping the site.
Reactions involving heavier species such as kaons or hyperons, among other, are not relevant since disk densities do not attain large values as their onset of kaons appears in most EoS models around three times saturation density $\sim 8\times10^{14}$  $\rm g/cm^3$. 
Our numerical  scheme calculates the absorption/emission rate as well as the energy-loss rates due to neutrinos. Specifically, we consider $\beta$-processes with electron-positron capture rate by nucleons $e^-+p\rightarrow n+\nu_e$, $e^+ +n\rightarrow p+\bar{\nu}_e$, plasmon decay $\gamma$ where quanta of electromagnetic field in a plasma, i.e. photons ($\gamma$) lead to neutrino-antineutrino pairs, $\gamma \rightarrow \nu_e + \bar{\nu}_e$, $\gamma \rightarrow \nu_x + \bar{\nu}_x$, where $x$ denotes generically the $\mu$ and $\tau$ flavours. We also consider electron-positron pair annihilation rate, $e^- + e^+\rightarrow \nu_e+\bar{\nu}_e$, $e^-+e^+\rightarrow \nu_x+\bar{\nu}_x$, and the nucleon-nucleon bremsstrahlung rate, $N + N \rightarrow N + N+ \nu +\bar \nu$. 

\begin{figure*}
    \includegraphics[width=0.48\linewidth]
   {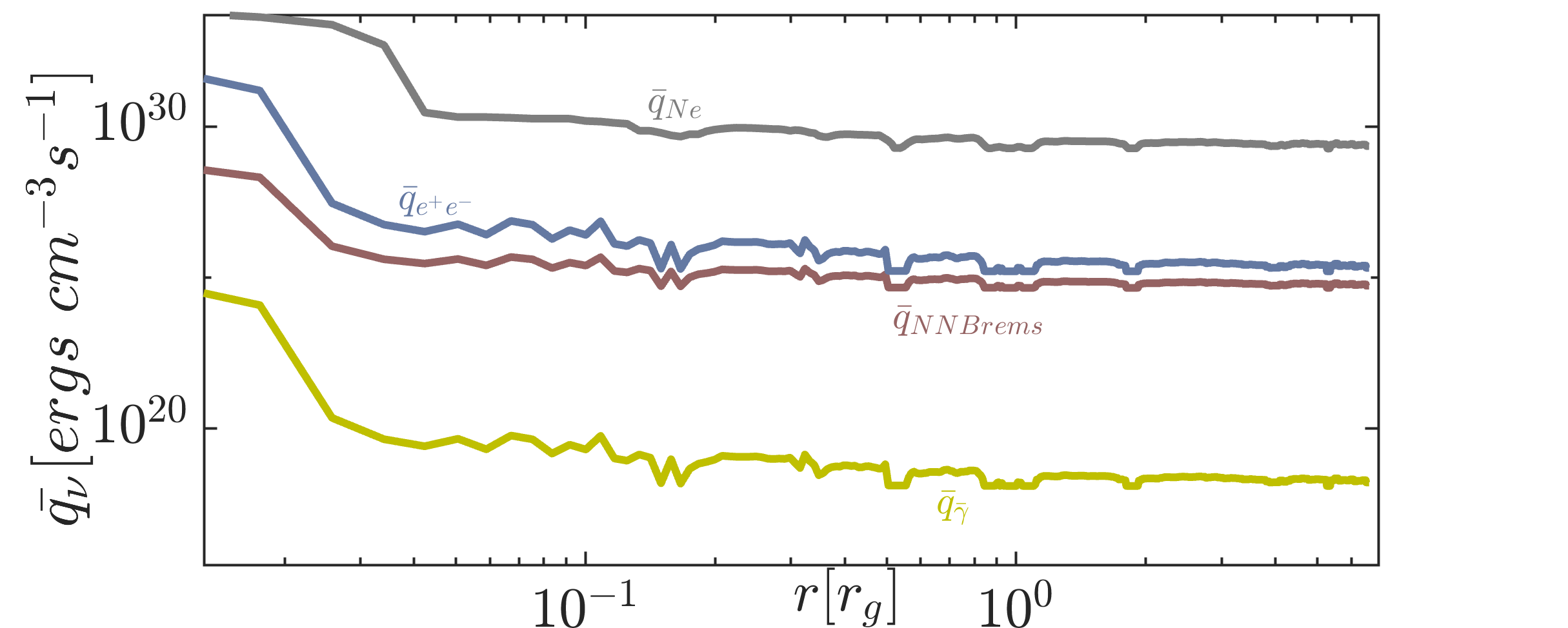} 
    \includegraphics[width=0.48\linewidth]
   {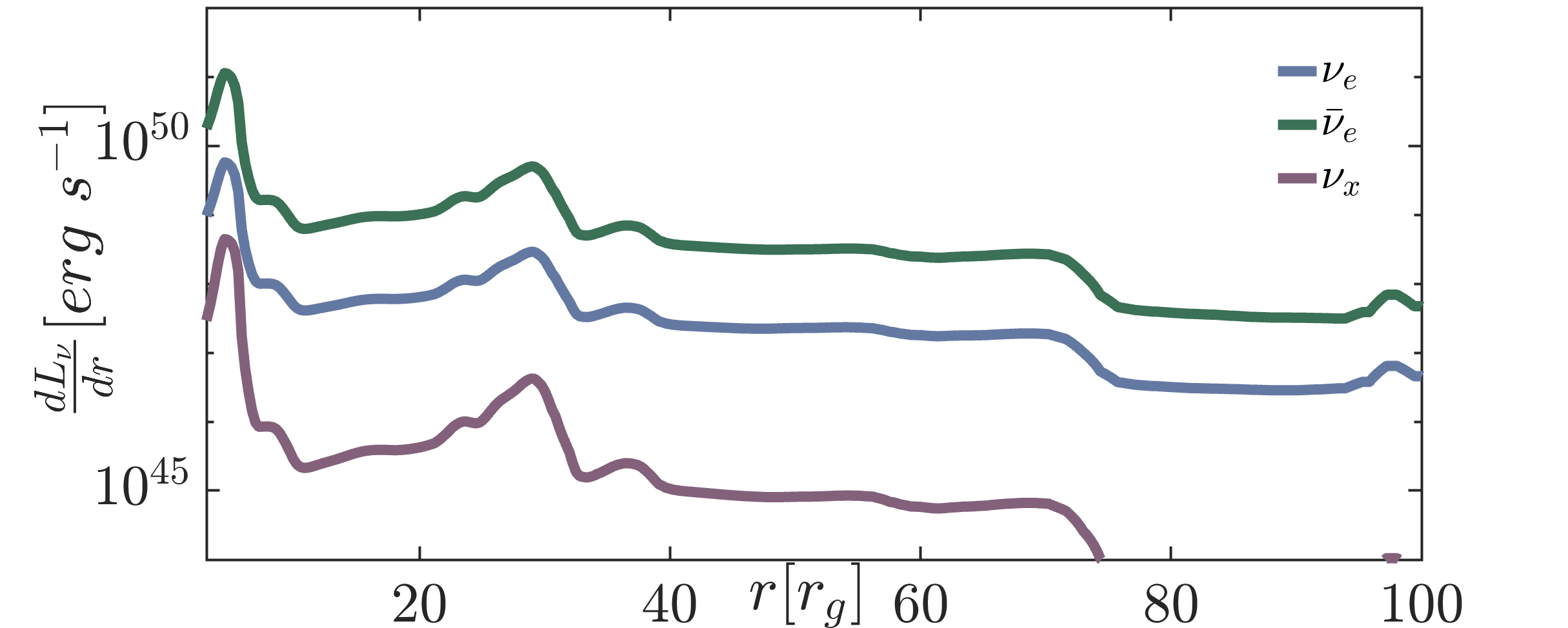}
        \caption{(Left) Neutrino emissivities per unit volume in the disk as a function of radial distance (in $r_g$ units) for the different processes considered i.e. Nucleon-electron (positron) $Ne$, electron-positron pair annihilation $e^+e^-$, NN bremstrahlung, ${NNBrems}$, and plasmon decay $(\gamma)$ up to $r=10r_g$. (Right) The luminosity due to the different neutrino species as a function of radius in $r_{g}$ units at $\rm t = 1\ ms$ after the NS collapse with initial $\rm B=0.5B_{16}$ where, $\rm B_{16} = 10^{16}G$.}
        \label{fig:Lum_rg}
\end{figure*}

In Fig. \ref{fig:Lum_rg}, top panel, we show the neutrino emissivities in the disk, $\bar q$ (per unit volume), as a function of radial distance (in $r_g$ units) for the different processes considered in our calculation, i.e. nucleon-electron (positron) $Ne$, electron-positron pair annihilation $e^+e^-$, NN bremstrahlung, ${NNBrems}$ and plasmon decay up to $r=10r_g$. In the bottom panel, we show the spherical differential neutrino luminosity (for e and $x\equiv\mu,\tau$ flavours) from the BH formation site up to the radial coordinate $r=100r_g$.
at $\rm t = 1\ ms$ after the collapse. 
The neutrino luminosity is large for all neutrino flavours close to the BH and quickly fades by 3 orders of magnitude in the external disk regions, this effect being more dramatic for $x$. One crucial process of neutrino absorption by nucleons (p,n), $\nu + n \rightarrow e^- + p$, provides the main channel for the scattering optical depth.


\begin{figure*}
    \begin{center}
        \includegraphics[width=0.49\linewidth]{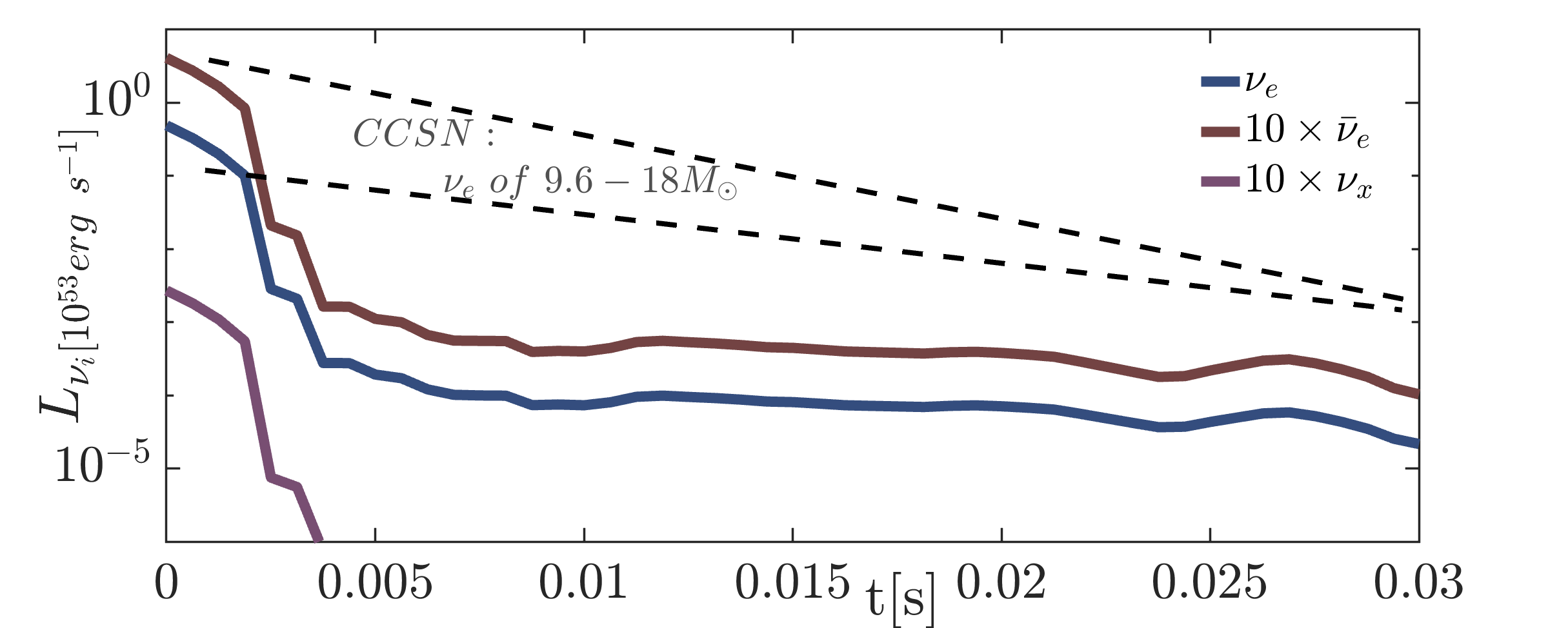}
        \includegraphics[width=0.49\linewidth]{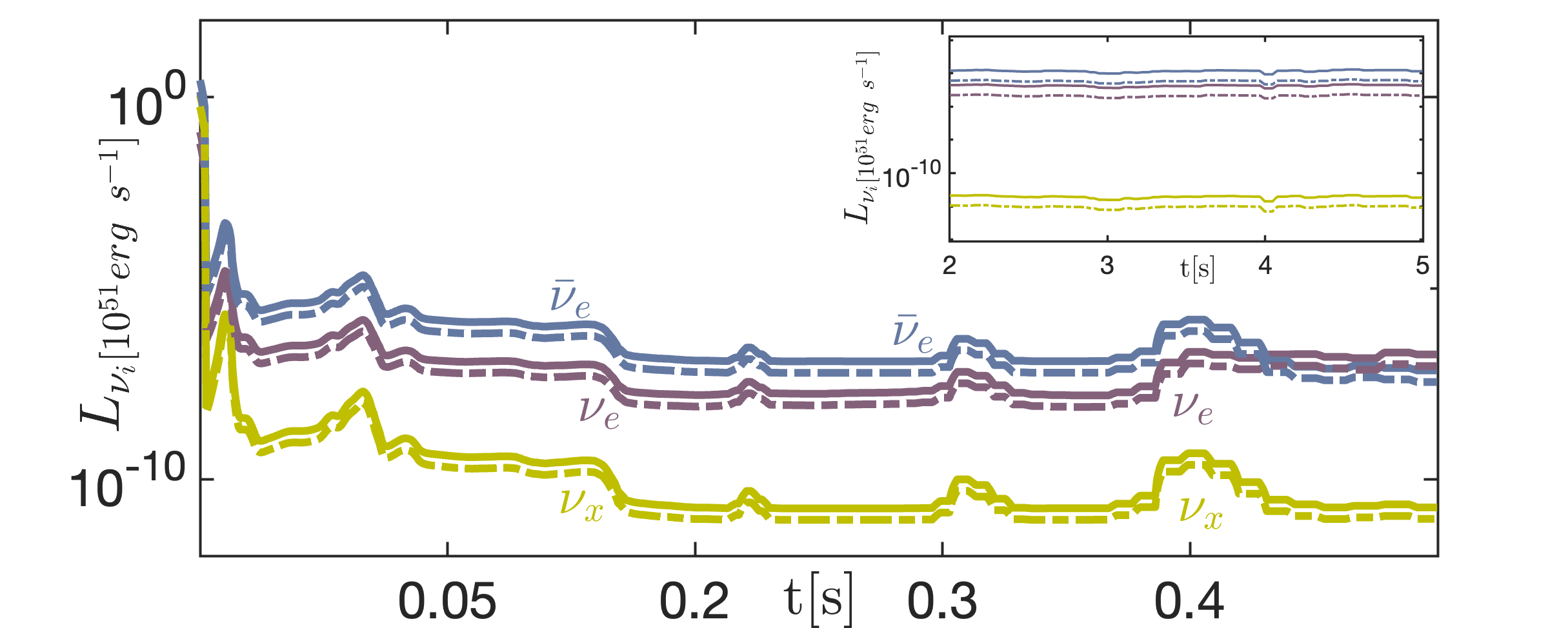}
         \caption{(Left) Time evolution of neutrino luminosities after collapse and formation of a BH $\rm M_{BH}= 1M_\odot$, for flavors $\nu_e$, $\bar \nu_{e}$, and $\nu_{x}$ (in units of $10^{53}\ erg\ s^{-1}$). We show the limiting curves for cases $[9.6,\rm 18] M_{\odot}$ CCSN \citep{ChakrabortyPRD+14} (Right) The neutrino luminosities for flavors $\nu_e$, $\bar \nu_{e}$, and $\nu_{x}$ (in units of $10^{51}\ erg\ s^{-1}$) evolution with time after bounce (in seconds) of the prompt collapse of NS to BH driven by DM capture. 
         The solid and dashed line elucidate a $\rm M_{BH}= 1M_\odot$ and a $\rm M_{BH}= 2 M_\odot$, respectively.}
        \label{fig:Lum_time_CCSN_NSBHprom}
    \end{center}
\end{figure*}
 Neutrino luminosities are shown in Fig. \ref{fig:Lum_time_CCSN_NSBHprom} (left) as a function of time from the BH formation site. They are compared to limiting cases (dashed lines) from a regular CCSN progenitor in the mass interval $[9.6,\rm 18]M_{\odot}$ \citep{ChakrabortyPRD+14}.
In more detail, we show specific cases for flavors $\nu_e$, $\bar \nu_{e}$, and $\nu_{x}$ (in units of $10^{51}\ erg\ s^{-1}$) evolving  with time after bounce (in seconds) of the prompt collapse for $1M_\odot$ (solid line) and $2M_\odot$ (dashed line) BH in right panel of Fig.\ref{fig:Lum_time_CCSN_NSBHprom}
 
Let us note that to evaluate  the neutrino luminosity, a gravitationally redshifted volume element integration of neutrino emissivities is involved \cite{Yakovlev&Pethick04} (the same applies to $L_\nu$ itself or T for local and distant observers). These $\mathcal{O}(1)$ factors due to curved space non-flat metric will hardly modify the large numerical values obtained roughly as $L_\nu\sim \int {\bar q}_{\nu} dV$, and the photon luminosity is 
obtained as $L_\gamma \sim 4 \pi R^2 \sigma T^4$ with $\sigma$ the Stefan-Boltzmann constant. For most of the emission cases analyzed, the neutrino luminosities associated with the NS implosion creating the BH will be systematically fainter than regular CCSN, thus for distant events a few neutrinos may reach detectors on Earth such as SuperKamiokande, ANTARES or IceCube. Although neutrinos from  catastrophic implosion events or binary mergers could  open a new window for multi-messenger research by looking for coincidence signals at times around GW detections, it seems technically very challenging, having yielded null results so far even for more massive, larger than $\sim 3\rm M_\odot$ resulting objects such as those from GW170817. Several searches have been performed for GW150914, GW151226, and GW170817 using a wide energy range from 3.5 MeV–100 PeV. The Superkamiokande analysis was performed within a time window of $\pm$500 s, as well as 14 days after the events finding only neutrino events compatible with the background\cite{Abe_2016,Abe_2018} within a distance of 260 kpc at $90\%$ C.L. Similar null results are reported from more recent data \cite{abbasi2023search,ANTARES:2023wcj}.

In the next section, we will discuss how this phenomenon is expected to be associated with both electromagnetic (EM) and GW signals. Even if we do not focus on the former here, see \cite{P_rez_Garc_a_2013}, the reason why some EM burst is expected boils down to the no-hair theorem and the fact that NS are magnetized nuclear objects, so adequate solutions with vanishing magnetic field must arise as the spacetime approaches that of a vacuum BH. Thus most of the EM energy is radiated away on dynamical timescales, giving rise to a transient signal.

\textit{Gravitational waves.—} As the NS accretes DM, the host star develops two matter spatial distributions inside, $\rho_1(r),\rho_2(r)$. As described, this may finally lead to the NS implosion. Apart from the expected multi-messengers under the form of photons and neutrinos, emission of GWs may take place in the catastrophic event. As is well known, no GWs are emitted from a spherically symmetric collapse. To our knowledge no numerical simulations have been performed on the likely off-center DM seeding and subsequent NS collapse. Due to the complex approach and for the sake of estimating the strength of the GW signal we will assume, in line with \cite{PhysRevD.103.044063}, that the collapse process is not fully axisymmetric and as a result, the $l=m=2$ and $l=2, m=0$ spherical harmonics components contribute to the quadrupole moment and waveform emitted. The former from the evolution of the quadrupole moment during the collapse process, and the 20 component comes from the collapsing, angular-averaged remnant, which is non-spherical due to its spin.

For an old NS collapse induced by accretion of a fermionic/bosonic DM component, there is a critical number or particles to trigger the NS transition to a $\sim M_\odot$ BH that can be estimated as
\small
\begin{equation}
N_{\chi,crit} \simeq 6 \times 10^{26}\left(\frac{10^{15} \mathrm{~g} / \mathrm{cm}^3}{\rho_c}\right)^{1 / 2}\left(\frac{T}{10^5 \mathrm{~K}}\right)^{3 / 2}\left(\frac{\mathrm{10^8 \,GeV}}{m_\chi}\right)^{5 / 2}
\end{equation}
\normalsize
where $\rho_c$ is the ordinary matter central density and we have assumed, as an example, the ADM particle mass $m_\chi\sim 10^8$ GeV, for additional parameter space see Fig.(1) in \cite{PhysRevLett.126.141105}. Note, however, that if the self-annihilating nature of DM candidate is assumed, a seeding mechanism may lead to collapse with smaller critical masses \cite{Perez-Garcia+10,Herrero+19}.

Assuming radial oscillations triggered in the collapse the GW signal strength in the BH formation is roughly estimated as \cite{kurita}
\begin{equation}
h \approx 2.8 \times 10^{-23}\left(\frac{d}{8 \,\mathrm{kpc}}\right)^{-1} \left(\frac{m_\chi}{10^8 \rm \,GeV}\right) \left(\frac{v_m}{0.1c}\right)^2 , \,\label{GW_h}
\end{equation}
where $d$ is the distance to the collapsing NS, typically clustering in the galactic center, and $v_m$ and $m_m$ denote the velocity and mass of the asymmetric component of the collapsing matter, respectively. Adopting typical velocity $v_m \approx 0.1c$ and assuming that fraction $f_\chi$ of the mass of the DM inner core contributes, typically $f_\chi \sim 0.01$, $m_m=$ $f_\chi \mathrm{m_\chi} N_{\chi,crit}$, contributes to the GW emission with strengths only accessible to the sensitivity of current detectors in the $0.1-1$ KHz. However, if the fraction of matter involved in the asymmetric collapse is much smaller or the DM candidate is not so massive, as predicted for bosonic candidates triggering efficiently the low mass BH formation, the strength may dramatically fall below current limits. In view of a similar result from other candidates, such as PBHs, in \cite{PhysRevLett.126.141105} an additional characteristic feature could lead to identify the transmuted origin of low mass black holes from the redshift dependence of their merger rate.
Given the fact that future missions such as Cosmic Explorer or the Einstein Telescope can only test strains $\rm Log (h) \gtrsim -25$ corresponding in this ADM setting to $m_\chi\sim 1$ TeV, lighter DM triggering collapse would not be feasible to detect these weak signals. Furthermore, some works \cite{PhysRevLett.128.111104} have shown promising results on sub-solar Mass BHs performing extreme mass-ratio inspirals (EMRIs) around a supermassive BH and SNR in the range of detectability by LISA or ET.

Regarding GW frequencies $f_{GW}$, an accurate determination is out of the scope of this work but their order of magnitude can be estimated from the oscillation phase as in \cite{kurita}, via the velocity associated to the asymmetric  matter component concentrated in regions of typical size $r\sim$ few cm. Thus $f_{GW}\sim v_m/r$ yields an expected range $[0.1,1]$ GHz. For the GHz band range, we need other technologies such as those discussed in \cite{FrancioliniPhysRevD.106.103520} and \cite{Aggarwal2021}. At this point, given the large uncertainties from the lack of reliable simulations, we must note that some analytical estimates \cite{PhysRevD.102.083004} find instead $\sim$kHz emission for other DM candidates, such as PBHs with larger masses and orbital radii when captured by NSs.


\textit {Conclusions and discussion.—} We have argued that accumulation of sufficient, for example asymmetric, DM inside a single NS may trigger the collapse into a low mass BH, as represented in the scenario depicted in Fig \ref{fig:spinning}, and that this releases a unique neutrino signal and  possibly also an ultrahigh-frequency GW signal in the $\sim$ GHz band. Such emissions are not accessible with current interferometer devices and could be detected via future experiments based on resonant cavities.
In addition to bosonic/fermionic ADM and self-annihilating candidates with feeble interactions with ordinary matter, there has been further recent discussion on PBH-NS accretion or collisional capture and the path of sink PBH material into the NS. 
Associated phenomena such as mechanisms for fast radio bursts (FRB) warm up the inner core. We do not address the source enigma of the FRB which is the reconnecting NS magnetic field \citep[see,][]{Petroff+19}. 
Subsequently, FRBs could occur during the collisions of NS with PBH pending released energy via gravitational drag. This PBH-NS collision is a probe of DM PBH-dominated galactic halos \citep{AbramowiczM+18}.
The light component, that is an imploding  solar mass NS,  including the EoS in relativistic and non-relativistic regimes\citep{Morras+23}, could play an important role in interpreting the trigger of a LIGO subsolar mass candidate. 
Simulations \citep{ZouZeHuang22} of PBH capture by both NS and strange stars show GW signal differences at kHz frequencies, potentially resembling our case of DM-induced collapse to a BH and suggesting that this might be a novel probe of the dense matter equation of state. Such collapsing NS are candidates for FRB \citep{AbramowiczM+18, KainulainenK+21}.

We have focused on the low mass BH fate of these NSs given the conditions discussed. The neutrino luminosity curve obtained from our calculation shows a unique and much fainter character than for regular CCSN. Strategies of detection based on coincidence signals from multi-messengers seem very challenging with current technologies although a constraining of the DM phase space may be at reach in the particular scenario of imploding NS considered here.


\begin{acknowledgments}
C. A. and M.A. P. G. acknowledge partial support from Junta de Castilla y Le\'on  SA096P20 and Spanish Ministry of Science  PID2019-107778GB-100 projects.  
\end{acknowledgments}

\bibliographystyle{apsrev}
\bibliography{apssamp}

\end{document}